\documentclass{aa}
\usepackage{graphicx}
\begin{document}

\newcommand{\GA}{\mbox{\raisebox{-0.6ex}{$\stackrel{\textstyle>}{\sim}$}}}
\newcommand{\LA}{\mbox{\raisebox{-0.6ex}{$\stackrel{\textstyle<}{\sim}$}}}

   \title{ROSAT/Chandra observations of a bright transient in M81}

   \subtitle{}

   \author{Kajal K. Ghosh
          \inst{1}\fnmsep\thanks{NRC Senior Research Associate}
          \and Douglas A. Swartz
          \inst{2}
          \and Allyn F. Tennant
          \inst{3}
          \and Kinwah Wu
          \inst{4}\fnmsep\thanks{ARC Australian Research Fellow}
              }

   \offprints{K. K. Ghosh}

   \institute{NRC, NASA Marshall Space Flight Center, SD-50,
              Huntsville, AL 35812, USA \\
              \email{kajal.ghosh@msfc.nasa.gov}
         \and USRA, NASA Marshall Space Flight Center, SD-50,
               Huntsville, AL 35812, USA \\
             \email{swartz@sdms.msfc.nasa.gov}
         \and Space Science Department,
             NASA Marshall Space Flight Center, SD-50,
               Huntsville, AL 35812, USA \\
            \email{allyn.tennant@msfc.nasa.gov}
         \and Mullard Space Science Laboratory,
              University College London, Holmbury St Mary,
              Surrey RH5 6NT, UK  \\
           and School of Physics A28, University of Sydney, NSW 2006, Australia \\
                \email{kw@mssl.ucl.ac.uk} }

   \date{
    }

   \abstract{
      We present a 10-year X-ray light curve
         and the spectra
         of a peculiar X-ray transient in the spiral galaxy M81.
      The source was below the detection limit of {\em ROSAT} PSPC
         before 1993,
         but it brightened substantially in 1993,
         with luminosities exceeding the Eddington limit
         of a 1.5-M$_\odot$ compact accretor.
      It then faded and was not firmly detected
         in the {\em ROSAT} HRI and PSPC observations after 1994.
      The {\em Chandra} image obtained in 2000 May, however,
         shows an X-ray source
         at its position within the instrumental uncertainties.
      The {\em Chandra} source is coincident with a star-like object
         in the Digitized-Sky-Survey.
      A {\em Hubble} image suggests that the optical object may be extended.
      While these three observations could be of the same object,
         which may be an X-ray binary containing a black-hole candidate,
         the possibility that
         the {\em ROSAT} and {\em Chandra} sources are
         two different objects in a dense stellar environment
         cannot be ruled out.
      The {\em Hubble} data suggests that the optical object
         may be a globular cluster yet to be identified.
   }

   \maketitle
%

\section{Introduction}

M81 (NGC 3031) is a nearby Sab spiral galaxy
   just beyond the Local Group.
It has a well defined central bulge;
   its two spiral arms are clearly marked
   by strings of bright young stars.

M81 was observed by {\em Einstein} in 1979
   and by {\em ROSAT} over the period 1991--1998.
The {\em Einstein} observations
   identified 10 point sources (including the nucleus)
   in the field (Fabbiano 1988).
There are 46 HRI and 69 PSPC sources
   found in the deep images obtained by {\em ROSAT}
   (Immler \& Wang 2001).
The {\em Chandra} observation was carried out in 2000.
A total of 97 sources were detected in the ACIS S3 image
   (Tennant et al.\ 2001),
   which has an area of $8'.3 \times 8'.3$
   centered at the nuclear region.
More sources are in the other S-chip images
   further away from the nucleus
   (Swartz et al.\ in preparation).

While a portion of the sources in the M81 field
   are foreground stars and background AGNs,
   the majority are believed to belong to the galaxy
   or its dwarf satellites
   (Tennant et al.\ 2001; Immler \& Wang 2001).
These X-ray sources form an inhomogeneous group,
   consisting of objects from canonical sources
   such as supernova remnants and X-ray binaries
   to exotic systems
   such as the very-soft sources discussed in Swartz et al.\ (2001).
Many sources are variable,
   and some are transients showing dramatic brightening
   for a brief duration ($\LA$1~yr)
   (see Immler \& Wang 2001; La Parola et al.\ 2001).

   \begin{figure*}
   \centering
   \includegraphics[angle=90,width=8.5cm]{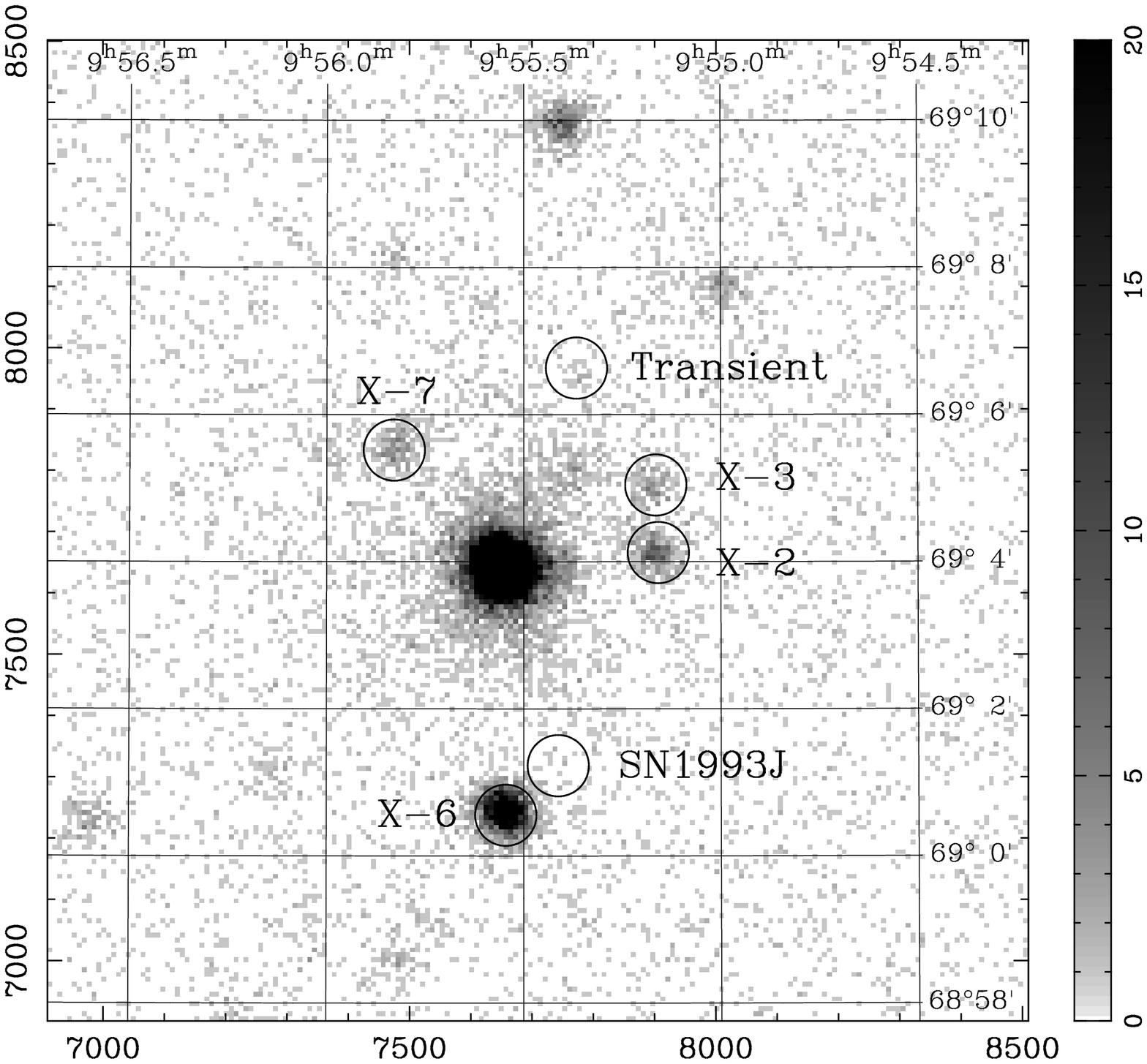}
   \includegraphics[angle=90,width=8.5cm]{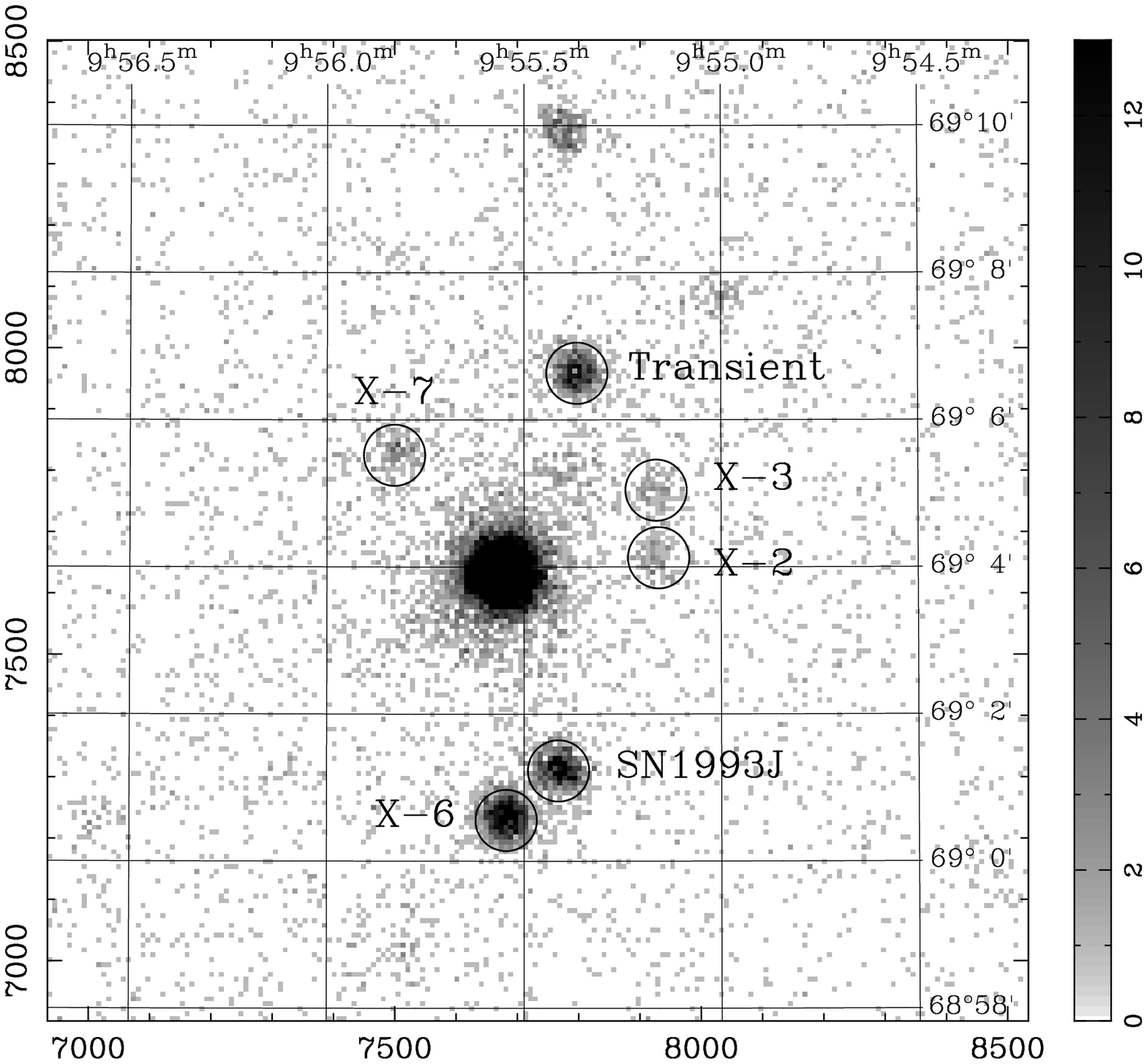}
   \caption{Comparison of two {\em ROSAT} PSPC observations in 1992 and 1993
      (left and right panels respectively).
      The bright nucleus is seen centered in the two images as well as
        several other sources in the field.
      One can see the brightening of the transient source P33
         and the supernova SN 1993J.
      The color table has be adjusted by the integration time so that
         sources of equal flux levels appear at similar gray levels.}
              \label{f:images}
    \end{figure*}

X-ray transients are also commonly found in our Galaxy.
Among the brightest are candidate black-hole X-ray binaries,
   whose peak luminosities may reach $10^{39}-10^{40}$~erg~s$^{-1}$
   during the outbursts,
   and certain neutron-star X-ray binaries (including the Type I bursters),
   which may be as bright as or even brighter than
   $2\times 10^{38}$~erg~s$^{-1}$,
   the Eddington limit of a 1.5-M$_\odot$ compact accretor.
These luminous X-ray binaries
  can be detected easily everywhere within the Galaxy
  when they are in active states.
(For a review of the properties of galactic X-ray binaries,
  see articles in Lewin, van Paradijs \& van den Heuvel 1995).
Transient X-ray binaries are found in globular clusters
  as well as in the Galactic bulge and disk. 
No candidate black hole X-ray
  binaries have been identified in Galactic globular clusters.
The bright globular cluster sources in our Galaxy
  tend to be Type I X-ray bursters
  (Verbunt \& van den Heuvel 1995). 
The most luminous globular cluster X-ray
  source in the Galaxy has a luminosity $\sim$5$\times 10^{37}$ erg~s$^{-1}$
  (Bloser et al. 2000).

Here, we report the analysis of the photometric and spectroscopic data
   of the peculiar transient source in the NW region of M81
   observed by {\em ROSAT} and {\em Chandra}.
The nature of the source and its implications are discussed.


\section{Observations and Data Analysis}

The M81 field was observed many times by {\em ROSAT} in the 1990s
  (Immler \& Wang 2001) including 9 PSPC observations
  spanning 1991 March to 1994 April
  and 11 HRI observations
  (from 1992 Oct to 1998 April).
We examined these data using a combination of FTOOLS and LEXTRCT, a locally
  supported software package.
{\em ROSAT} PSPC images obtained in 1992 and in 1993 are shown in
  the left and right panels of Figure~\ref{f:images}, respectively.
The transient as well as SN~1993J are both seen in the 1993 data.
In the 1992 data, there is only a very slight excess at the transient's
  position that is significant at the 2.4~$\sigma$ level.
To derive an accurate position, 
we adjusted the coordinates in each {\em ROSAT}
image
to align the nucleus to be at the radio position
  $9^{\rm h}55^{\rm m}33.17^{\rm s}$, +69~03~55.1 (Ma et~al. 1998). 
This
  adjustment was typically $\sim$5 arcsec. No rotation was applied.
With this offset, the locations of SN~1993J and the bright {\em Einstein} sources
  X-2, X-3 and X-6
  are all consistent with their {\em Chandra} positions.
The transient is listed as source P33 in Immler \& Wang (2001)
  at ($9^{\rm h}55^{\rm m}22.7^{\rm s}$, +69~06~31).
In our adjusted coordinates, its position
  is ($9^{\rm h}55^{\rm m}21.96^{\rm s}$, +69~06~38.0)
  with an uncertainty of 3.5~arcsec.

The M81 field was observed for 50~ks by {\em Chandra} on 2000 May 7
  with the nucleus centered on the ACIS S3 chip.
The 
{\em Chandra} image, 
data-analysis,
  and general properties of the sources found in the S3 image
  were presented in Tennant et al.\ (2001).
The transient is less than 1~arcsec
  from a {\em Chandra} source at a position of
  ($9^{\rm h}55^{\rm m}21.95^{\rm s}$, +69~06~37.8).
The source is off-axis in the {\em Chandra} data,
  appearing roughly elliptical in shape,
  with Gaussian widths $\sigma=$ 2 and 1 arcsec in the two dimensions on the sky.

The {\em Chandra} position is less than 1 arcsec from the object
listed as ID~50777 by Perelmuter \& Racine (1995) who report
   its brightness and colours are $V=16.87$, $B-V=0.97$, and $V-R=0.56$.

The region around object ID~50777 was observed with
  {\em Hubble} WFPC2 on 1998 April 17.
The three images taken with the F450W filter,
  which roughly correspond to the V band,
  were combined using a median filter to remove cosmic ray tracks.
The final image was visually inspected and
  in particular object ID~50777 was compared to object ID~50658;
  another bright star-like object with V=15.91 (Perelmuter \& Racine, 1995).
Object ID~50658 shows diffraction spikes and is saturated in the core with
  indication of some charge leakage during read out.
Object ID~50777 does not show diffraction spikes, nor does it appear to be
  saturated in the core.
The background around both objects is partially resolved into stars and
  no attempt was made to remove these very faint stellar images.

\begin{figure}
   \centering
   \includegraphics[angle=90,width=8.5cm]{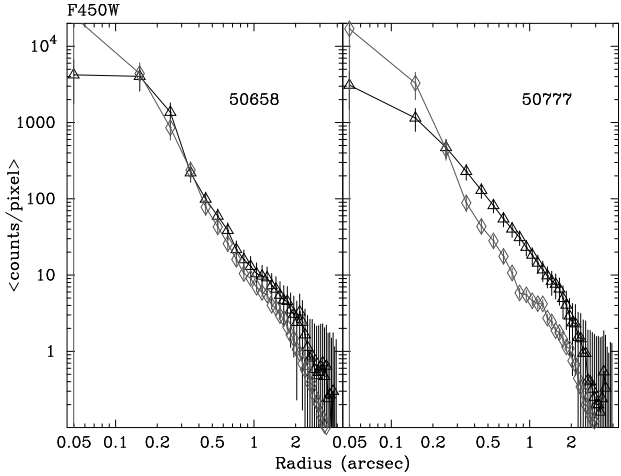}
   \caption{The radial profile of two objects
       in the {\em Hubble} WFPC2 data with the F450W filter.
     The profile from the data are shown in black triangles.
     Tiny Tim model PSFs normalized to contain the same number of
       counts are shown with gray diamonds.
     The error bars are based on the scatter and thus represent the
      azimuth variation in the PSF as well as
      the variations in the background due to faint stars.
       }
              \label{f:hst}%
    \end{figure}

The Tiny Tim software package (Krist \& Hook 1999) was
  used to compute model Point Spread Functions (PSFs) for the two locations.
The radial profiles of the two sources
  as well as the profiles derived from the Tiny Tim model PSFs
  are shown in Figure~\ref{f:hst}.
To normalize the data we subtracted the average level in the outer three
  annuli(from roughly 3.7 to 4.0 arcsec).
To normalize the Tiny Tim models we multiplied by the total number
  of counts detected in the data.
As can be seen the PSF of ID~50658 is saturated in the core with an excess
  at 0.25 arcsec where the charge is finally read out.
At larger radii, the model PSF approximates the data fairly accurately.
The model PSF poorly approximates the data for object ID~50777.
The radial profile of ID~50777 is very smooth and does not show the central
  core nor the excess at 1-2 arcsec (mainly due to the diffraction spikes).
Thus, ID~50777 appears extended in the {\em Hubble} data.
Not shown in the figure is the PSF for object ID~50696 (Perelmuter \& Racine 1995)
  which is also in the same {\em Hubble} field and was confirmed
  by Perelmuter, Brodie \& Huchra (1995) as a globular cluster in M81.
The radial profile of ID~50696 shows an extension very similar to that of object ID~50777.

We conclude that object ID~50777 is extended in the {\em Hubble} data and
  is very likely a bright globular cluster in M81.
The half light radius in the {\em Hubble} data is 2.4 pixels.
At the distance of M81 this corresponds to 4.4~pc.
If we remove the estimated {\em Hubble} PSF then the half light radius
  would be about 3~pc which is a fairly typical size for globulars
  in our galaxy ({\em e.g.},\ Ashman \& Zepf 1998).

\section{Brightness Variation and Spectroscopic Properties}

The 10-year X-ray light curve of the transient source is shown in Figure~3.
The source was below the {\em ROSAT} HRI equivalent count rate
  of $\sim 3 \times 10^{-4}$~cts~s$^{-1}$
  up to 1993 May in both the PSPC and HRI images.
Its count rate, however, reached about $1.5\times 10^{-2}$~cts~s$^{-1}$
  during the flare (brightening) around 1993 September$-$November.
It then faded within a 5-month period
  and was marginally detected ($< 2.5 \sigma$) in the subsequent
  {\em ROSAT} observations
  at the brightness level of $\sim (2-3) \times 10^{-4}$~cts~s$^{-1}$.

The equivalent {\em ROSAT} HRI count rate of the {\em Chandra} source
  is $(2.6\pm0.2) \times 10^{-4}$~cts~s$^{-1}$,
  consistent with the `quiescent' brightness of the source P33
  seen in the {\em ROSAT} observations.

We extracted the {\em ROSAT} PSPC spectra of the source
  during the brightening in 1993
  and grouped them into spectral bins
  containing a minimum of 20 counts.
The spectra are fitted using XSPEC (Arnaud, 1996) to simple functions.
The best fitting parameter values
  as well as the 90 percent confidence single parameter uncertainties
  are shown in Table 1.
In all cases, the absorption column density is consistent with
  the line-of-sight Galactic value
  ($n_{\rm H} = 4 \times 10^{20}$~cm$^{-2}$, Stark et al.\ 1992).
The power-law and bremsstrahlung models provide equally acceptable fits.
For the black-body model,
  the best fitting absorption column density is consistent with zero,
  and therefore, we fixed its value to the Galactic value.
Based on the value of $\chi^2$ we cannot rule out a black-body model
  although it does not fit as well.
When we add the two {\em ROSAT} data sets,
  the black-body model is still the worst model,
  but it cannot be ruled out at the 3-$\sigma$ level.

Assuming that the source is within M81
  and the distance to the galaxy is 3.6~Mpc (Freeman et al. 1994),
  the black-body model implies an unabsorbed luminosity
  $\sim 6.7\times 10^{38}$~erg~s$^{-1}$ in the 0.2-2.4~keV band,
  and a negligible intrinsic absorption of the source
  in comparison with the line-of-sight Galactic absorption.
If the flare spectrum is a power law then the luminosity
  in the {\em ROSAT} band is $\sim$$8.5 \times 10^{38}$~erg~s$^{-1}$.
Clearly if the power law extends to higher energies, then
  the X-ray luminosity could be significantly above $10^{39}$~erg~s$^{-1}$.

The {\em Chandra} data show that
  the source had a power-law spectrum in `quiescence'.
The best fit parameters are listed in Table 1.
The inferred unabsorbed luminosity is $1.7 \times 10^{37}$~erg~s$^{-1}$
  in the 0.5-8~keV band
  and, in the {\em ROSAT} bandpass, it is roughly 60 times fainter than
  that of the flare luminosity in late 1993.
There is a suggestion that
  the {\em Chandra} spectrum is steeper than the {\em ROSAT} spectra;
  however we note that $\Gamma \sim 1.6$ is consistent with both data sets.

   \begin{figure*}
   \centering
   \includegraphics[angle=90,width=17cm]{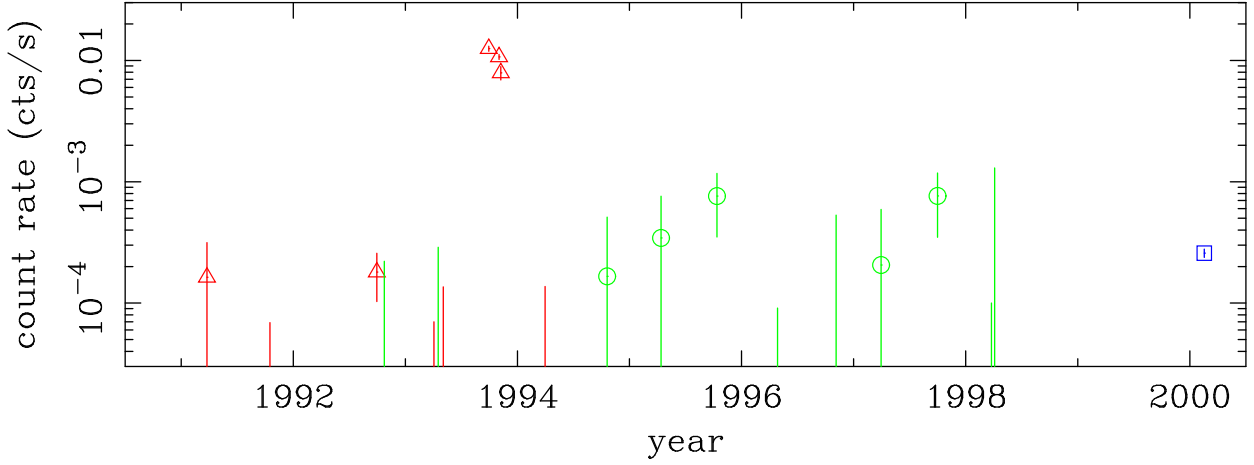}
   \caption{The light curve of the transient between 1991 and 2000.
      The count rates are converted to equivalent {\em ROSAT} HRI count rates,
         assuming that the spectrum is a power law
         with $\Gamma = 1.6$ and $n_{\rm H} = 4.0 \times 10^{20}$~cm$^{-2}$.
      Data before 2000 are the {\em ROSAT} observations:
         triangles correspond to PSPC observations
         and circles correspond to HRI observations.
      The {\em Chandra} observation in 2000 May
         is represented by a square.}
              \label{Figure 3}
    \end{figure*}

   \begin{table*}
      \caption[]{Spectral fit to the {\em ROSAT} and {\em Chandra} observations}
         \label{Table 1}
     $$
         \begin{array}{p{0.13\linewidth}p{0.13\linewidth}lcccc}
            \hline
            \noalign{\smallskip}
    observation & date & model & n_{\rm H}~(10^{20}~{\rm cm}^{-2})
          &  \Gamma & T (keV) & \chi^2/dof  \\
            \noalign{\smallskip}
            \hline
            \noalign{\smallskip}
   ROSAT & 1993~Sep~29
     &  phabs*po    & 4.4^{+2.7}_{-1.8} & 1.30^{+0.37}_{-0.36} &  & 19.6/22 \\
     & & phabs*bb   & 4.0~(fixed) & & 0.33^{+0.05}_{-0.33}    & 31.6/23 \\
     & & phabs*brem & 4.3^{+2.1}_{-1.2}& & 15.5^{+\infty}_{-10.1} & 19.6/22 \\
   ROSAT &  1993~Nov~1
     &  phabs*po    & 6.3^{+12.8}_{-3.2} & 1.55^{+0.95}_{-0.45} &  & 19.8/18 \\
     & & phabs*bb   & 4.0~(fixed) &  & 0.32^{+0.05}_{-0.32} & 22.5/19 \\
     & & phabs*brem & 5.6^{+7.1}_{-2.4}& & 4.7^{+\infty}_{-2.3} & 19.4/18 \\
   ROSAT & coadd~Sep/Nov
     &  phabs*po    & 4.8^{+2.6}_{-1.6} & 1.37^{+0.29}_{-0.27} &  & 46.0/43 \\
     & & phabs*bb   & 4.0~(fixed) & & 0.32^{+0.03}_{-0.02}  & 61.5/44 \\
     & & phabs*brem & 4.6^{+1.9}_{-1.0}& & 9.5^{+\infty}_{-3.9} & 45.6/43 \\
   Chandra  & 2000~May~7
     &  phabs*po  & 12.4^{+17.7}_{-12.4} & 1.88^{+0.60}_{-0.55} &  & 2.3/4 \\
     & & phabs*bb  & 4.0~(fixed) & & 0.47^{+0.10}_{-0.47}   & 10.6/5 \\
     & & phabs*brem & 7.6^{+3.9}_{-7.6}& & 4.5^{+\infty}_{-0.9} & 2.4/4 \\
            \noalign{\smallskip}
            \hline
         \end{array}
     $$
   \end{table*}

   \begin{figure*}
   \centering
   \includegraphics[angle=0,width=8.5cm]{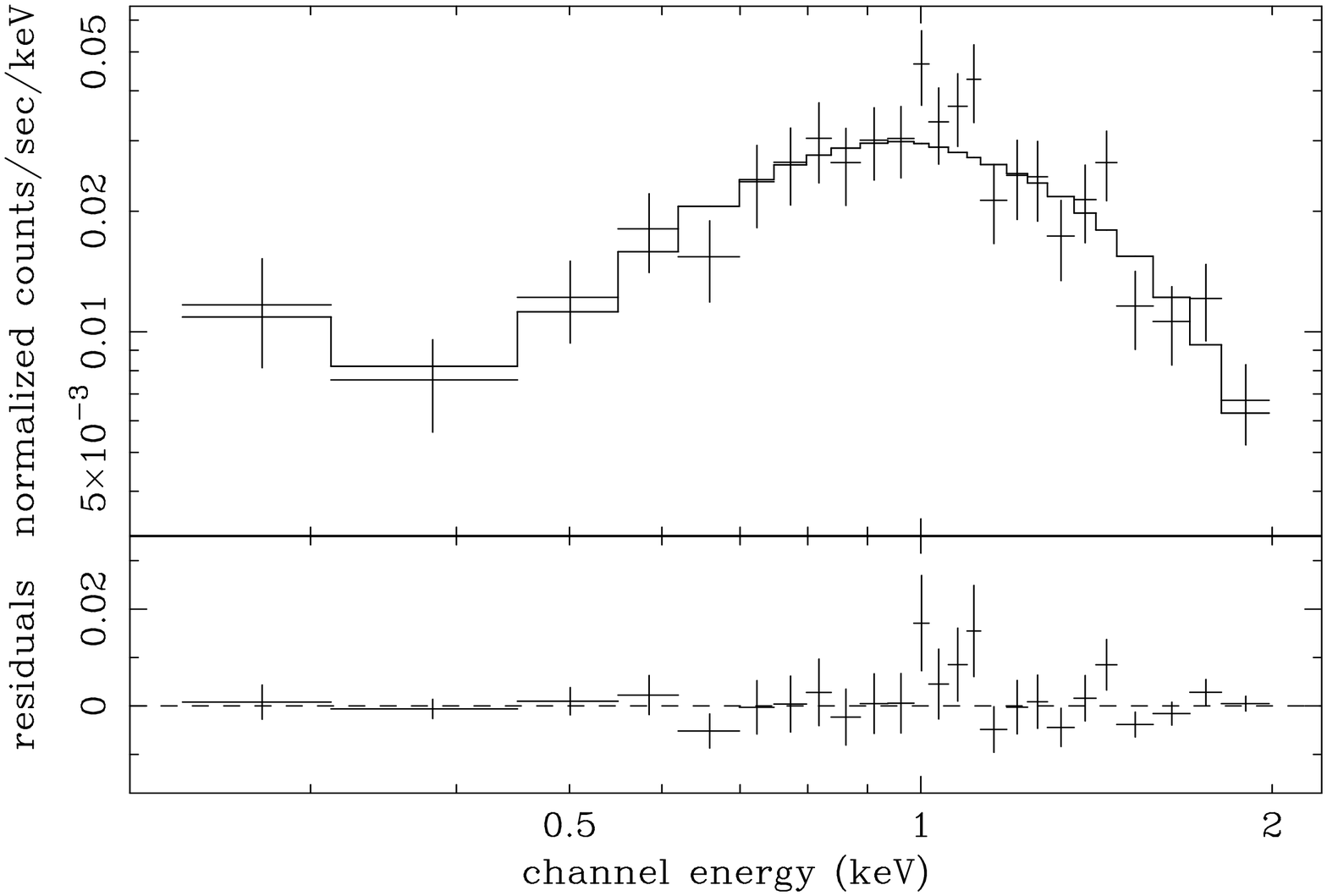}
   \includegraphics[angle=0,width=8.5cm]{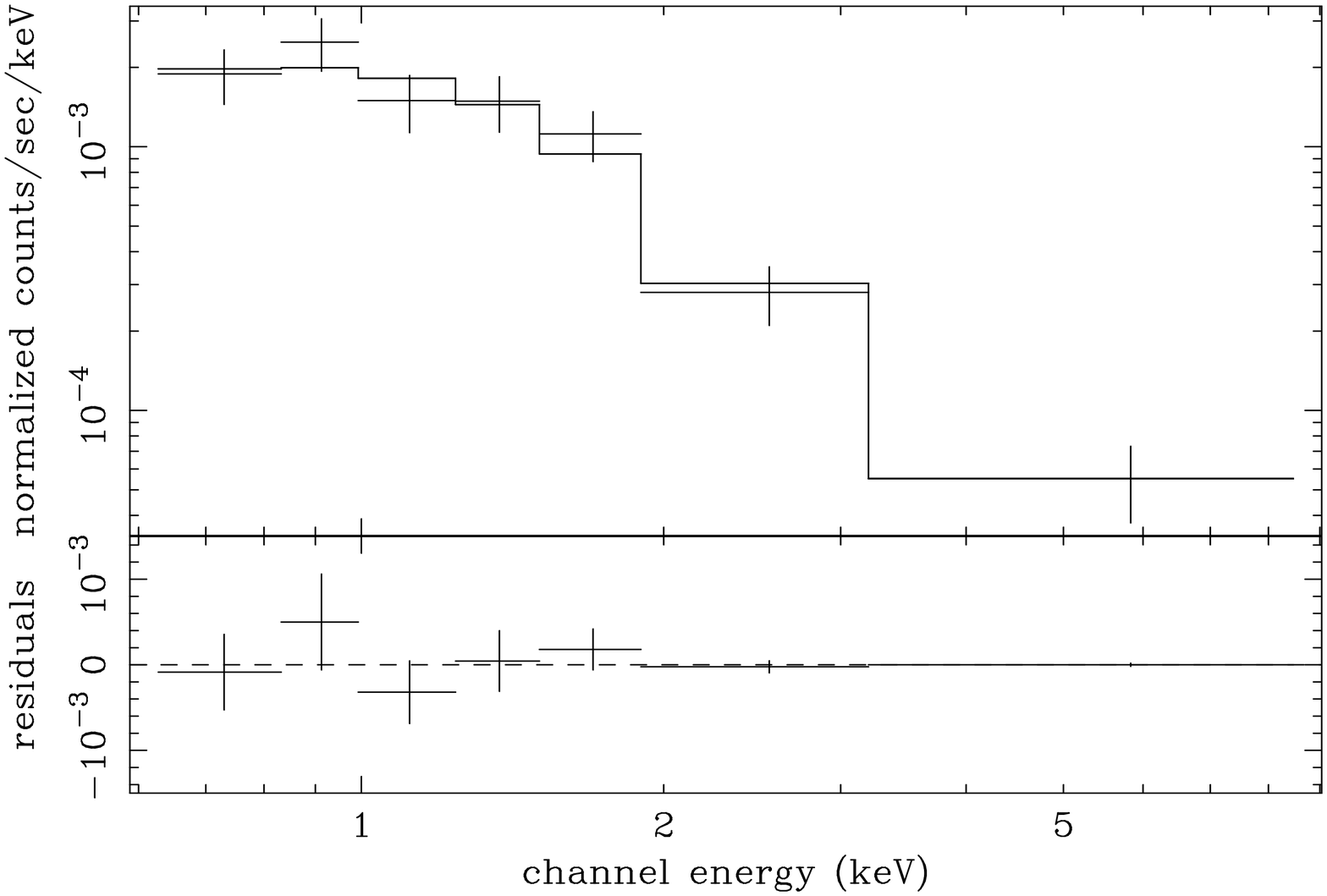}
   \caption{Fits to the {\em ROSAT} and {\em Chandra} spectra of the source
      (left and right panel respectively).
    The {\em ROSAT} spectrum was obtained on 1993 September 29
      when the source was bright.
      The model of the fit is (phabs*po) and the parameter values
      are listed in Table 1.
    The source was relatively faint
         during the {\em Chandra} observation on 2000 May 7.
      The spectrum is fit by a model (phabs*po) with the parameters listed
      in Table 1.
         }
              \label{Figure 4}
    \end{figure*}


\section{Nature of the Source}

The fact that the flaring spectra
  obtained by {\em ROSAT} PSPC in late 1993
  are equally well fit by a number of simple models implies that
  the exact nature of the source is still open to investigation.
The low absorbing column does tend to rule out a high mass companion,
  since such systems typically have strong stellar winds and
  hence large intrinsic columns.
Here we briefly discuss several possibilities.
We begin by assuming a single source is responsible for all the observed
  X-ray emission and describe possible scenarios for objects outside
  (\S~\ref{s:galactic}) or within (\S~\ref{s:m81xrb}) M81.
We then consider (\S~\ref{s:gcxrb}) the possibility that two or
  more X-ray sources are present along the same line of sight.

\subsection{Systems outside M81} \label{s:galactic}

If we assume the X-rays come from a star within our Galaxy
  and that object ID~50777 in Perelmuter \& Racine (1995) is its counterpart,
  the colours suggest an early K spectral type.
If the star is on the main sequence, the optical brightness implies a distance of 1.5 kpc.
We first consider the possibility that
  the X-rays are due to coronal emission.
For stellar coronae,
  the ratio of the X-ray flux to the optical flux
  is expected to be in the range $10^{-3}$ to $10^{-5}$
  for F- to M-type stars (Topka, et al. 1982)
  whereas we find $F_{\rm x}/F_{\rm opt} \approx 0.14$.
The hardness of the {\em Chandra} spectrum
  is also inconsistent with typical coronal emission.
Therefore, coronal emission from a late-type main-sequence star is unlikely.

If a distance of 1.5~kpc is assumed,
  then the flare luminosity of the source would be $\sim 10^{32}$~erg~s$^{-1}$.
Typical X-ray luminosities (above 0.5 keV)
  of magnetic cataclysmic variables (mCVs) ---
  which are close binaries consisting of a magnetic white dwarf
  and a late K or M main-sequence star ---
  are about $10^{31}-10^{32}$~erg~s$^{-1}$
  (see Cropper 1990 and references therein).
MCVs are also known to show transitions between high and low states
  (Cropper 1990, Warner 1995).
Thus we are unable to rule out a foreground mCV based
  on the X-ray observations alone.
However, the optical source should not appear extended as it does
  in the {\em Hubble} image.
The issue could be resolved if an optical image taken during the X-ray high
  state were available as mCVs show brightening in both the X-ray and optical
  bands during high states.

The extended emission could be accounted for if the object were a background
  AGN.
AGNs are not known to show such large amplitude X-ray variabiltiy
  on timescales of a few months as seen.
In principle tidal disruption of a star by the black hole at the center
  of a galaxy could produce a large amplitude ($10^3$ to $10^4$) flare
  but the expected timescale is of order a year or more (Ulmer 1999).
Furthermore, if the object is behind M81, then the absorbing column
  would come from not only our own galaxy, but also
  the column through the entire disk of M81 as well as any intrinsic column.
The low column measured from the X-ray data makes this possibility unlikely.
An optical measurement of spectral redshift would provide conclusive evidence.

If the optical source is unrelated to the X-ray source,
  then the true optical counterpart would be
  a fainter and presumably more distant object.
The most common X-ray source is an interacting binary.
X-ray binaries in their active states
  have luminosities in the range $10^{36}-10^{38}$~erg~s$^{-1}$
  ({\em e.g.}\ White, Nagase \& Parmar 1995).
But, the {\em ROSAT} count rate during the flare in 1993
  implies a luminosity of only $\sim 5 \times 10^{35}$~erg~s$^{-1}$
  in the 0.2$-$2.4~keV band even at a distance as great as
100~kpc, which would place it in the outer part of the Galactic halo.
If the source is an X-ray binary within the Galaxy, unrelated to the observed optical source,
  it would be one of the fainter systems.

\subsection{Systems within M81} \label{s:m81xrb}

If we assume that the optical object is in M81 and is related
  to the X-ray source, then it is far too bright to be
  an isolated X-ray binary in M81.
It is also too bright to be an isolated supernova remnant
  which could potentially account for the optical extension,
  but not the X-ray variability.

The optical object may be
  a previously unidentified (and yet to be confirmed) globular cluster in M81
  harbouring the X-ray source.
Galaxies similar to M81 could contain of order 200 globular clusters
  (Perelmuter \& Racine 1995).
To date, only 25 globular clusters in M81 have been confirmed
  (Perelmuter, Brodie \& Huchra 1995),
  and only 4 of these are within the viewing field of ACIS~S3.
The colours reported by Perelmuter \& Racine (1995) for ID~50777 are
  within their expectations for M81 globular cluster candidates.
As shown in section 2, object ID~50777 has a half light radius of
  about 0.2 arcsec which would be impossible to detect from the ground,
  although the outer halo could be.
The V magnitude of 16.9 would
  imply an absolute magnitude of $M_{\rm V} \approx -10.9$
  which would make it brighter than any Milky Way cluster and
  similar to the absolute magnitude
  of the brightest globular cluster in the Andromeda galaxy M31
  (see the review by Ashman \& Zepf 1998).
Thus the brightness and small size of ID~50777 would account for it
  being missed in previous searches for globular clusters.

X-ray transients are known to occur in Milky Way globulars.
Typical globular cluster sources in our galaxy have luminosities ranging up to
  $\sim 5\times 10^{37}$~erg~s$^{-1}$ and none are black hole candidates.
The luminosity of the {\em Chandra} source
  is consistent with that of bright X-ray sources found
  in Galactic globular clusters
  (Verbunt \& van den Heuvel 1995, Sidoli, et al. 2001).
Both the spectra of the {\em ROSAT} observations in late 1993
  and the {\em Chandra} observation are well fit by
  a power-law with a photon index about 1.6.
This value is consistent with the power-law component
  seen in the spectra of X-ray binaries.
The {\em ROSAT} count rates together with the distance to M81
  imply the flare luminosity of the source above
  the Eddington luminosity of X-ray binaries
  containing an accreting neutron star.
Di~Stefano et~al. (2001) report {\em Chandra} observations
  of M31 that show that the majority of the most luminous X-ray sources in
  M31 are in globular clusters and that the highest sustained luminosity
  observed exceeds $2 \times 10^{38}$~erg~s$^{-1}$.
We also note that some peculiar Galactic neutron-star binaries
  ({\em e.g.} Cir X-1, which is believed to have a very eccentric orbit)
  have super-Eddington luminosities
  (up to three times the Eddington limit of a 1.5-M$_\odot$ accretor)
  over a quite substantial portion of their orbital cycles
  ({\em e.g.}, Shirey et~al. 1996).
Thus the possibility of an extremely bright neutron-star X-ray binary cannot be
  easily eliminated.

The high luminosity makes the source a candidate black-hole binary.
From the {\em Chandra} observation
  we obtain a `quiescence' luminosity of $1.7 \times 10^{37}$~erg~s$^{-1}$.
The ratio of the flare luminosity to the quiescence luminosity
  is therefore $\sim$60.
This is significantly lower than those of the Galactic black-hole binaries
  GS 2000+251, Nova Mus 91, A0620-00, and V404 Cyg
  whose observed ratios ($10^{6}$ to $10^7$, Garcia et al. 2001) are
  in agreement with Advection-Dominated-Accretion-Flow (ADAF)
  model predictions (Narayan, Garcia \& McClintock 1997).


\subsection{Multiple X-ray sources in a M81 globular cluster}
\label{s:gcxrb}

Although the X-ray temporal behaviour suggests an X-ray transient,
  the peak luminosity is difficult to explain with a neutron star.
On the other hand the brightness of Chandra measurement is inconsistent
  with the currently popular ADAF model.
This leads to the possibility that
  the strong source seen by {\em ROSAT} in 1993
  and the fainter source observed by {\em Chandra} in 2000
  are two different objects.
If we assume that each detected X-ray source occupies an area
  of the detector out to the 90\% radius of the {\em Chandra} PSF then
  the 97 X-ray sources reported in Tennant et al. (2001) would occupy
  $\sim$2\% of all the pixels on the ACIS S3 detector.
Thus the probability of a randomly-located source
  appearing within a detected ACIS source region is 2\%.
Of course, assuming X-ray sources are more likely to be found in
  dense stellar environments, such as the cores of globular clusters,
  then this probability should be considered a lower limit.

Thus, object ID~50777 may be a massive
  ({\em i.e.,}\ luminous and compact) globular cluster in M81
  containing the {\em ROSAT} flare source P33
  and the fainter {\em Chandra} source.
The {\em Chandra} source is presumably 
one or more persistent X-ray binaries
while
  the brightness of the {\em ROSAT} source in 1993 suggests
  a black-hole binary in the core.

If systems displaying similar X-ray characteristics can be identified elsewhere,
however, the multiple-source hypothesis becomes less viable and a single-source
explanation is to be favored.
We note that the {\em Einstein} source X-2 lies very close to a bright
  (V=17.29) point-like object listed as ID~50548 in Perelmuter \& Racine (1995).
During the {\em Chandra} observation, X-2 had faded dramatically relative to
its {\em Einstein}-measured brightness.
To our knowledge, however, X-2 has never exceeded an X-ray luminosity of
  $\sim 2 \times 10^{38}$~erg~s$^{-1}$.
Optical followups of ID~50548 would also be of interest.
The recently discovered transient source (Steinle, Dennerl \& Englhauser 2000)
  in the direction of galaxy Cen~A may also
  display characteristics similar to those reported here.
If the source is in Cen~A, then it would have had a peak X-ray
  luminosity of $3 \times 10^{39}$~erg~s$^{-1}$ and
  it also appears to be associated with a weak {\em Chandra} source.

{Roughly
  10\% of globular clusters are expected to contain accreting neutron stars
  detectable above the {\em Chandra} threshold and galaxies like M81 typically
  contain of order 200 globular clusters (Perelmuter \& Racine 1995). However,
  there are only 25 confirmed globular clusters in M81 (Perelmuter, Brodie, \&
  Huchra, 1995) and only 10 within the {\em Chandra}-observed field. None of
  these are coincident with X-ray sources detected by {\em Chandra}.}


\section{Summary}


The transient X-ray source brightened to
$\sim$$7 \times 10^{38}$~erg~s$^{-1}$,
during {\em ROSAT} observations spanning 1993 Sep 29 to
1993 Nov 7 and was only 12--75 times fainter during the rest of the
10-year observing period including
observations taken 5 months before and after the bright phase.
Spectra obtained during both the
high and low brightness states can be equally well fit by either
power law models with $\Gamma \sim 1.6$ and by
thermal bremsstrahlung models with temperatures exceeding 4~keV.
In all cases the data are consistent with only a Galactic interstellar
absorption column.
Blackbody models provide a substantially poorer fit.
The bright optical counterpart to the transient source is extended
with a half-light radius of $\sim$0.2~arcsec and with luminosity and
colours consistent with a massive globular cluster in M81.

The only viable foreground object candidate for this source is a mCV.
A Galactic mCV would have a flaring luminosity (at a distance $\sim
1.5$~kpc) and spectral shape similar to that observed.
The optical extension and low state X-ray luminosity of order
  $\sim 2 \times 10^{30}$~erg~s$^{-1}$ argue strongly against this hypothesis.
A background AGN is unlikely based on the short X-ray ``on-time''
of the source and on the low column density derived from the spectral
fits.

An X-ray binary at the distance of M81 is suggested by the shape of
the observed power law spectra yet the luminosities observed
make this scenario difficult to reconcile
with canonical Galactic objects in this class:
The high luminosity during the flare
exceeds that of an accreting neutron star while the small ratio of flare
luminosity to quiescent luminosity argues against a black-hole system.
This can be resolved if the X-rays come from two objects,
  a black hole that produced the flare and
  a neutron star to account for the {\em Chandra} flux.
Though a rare superposition may occur, the probability is
  greater if the sources are located
  within a single massive globular cluster in M81. 
{If this is the case, then
  this is the first black hole candidate in a globular cluster. In any case, the
  peak X-ray luminosity exceeds that of any globular cluster X-ray source in the
  Galaxy and rivals that of the brightest M31 globular cluster X-ray source.}

Further optical observations are encouraged in order to confirm the
nature of the optical counterpart and to monitor changes in optical
brightness if the source turns out to be a foreground object.
An optical redshift measurement would strictly limit the AGN hypothesis.
The companion star of a foreground mCV should show distinct emission
lines unlike the absorption line spectrum typical of globular clusters.



\begin{acknowledgements}

K. W. thanks Martin Weisskopf for funding his visits to MSFC. Support for this
research was provided in part by NASA/{\em Chandra} grant GO0-1058X.

\end{acknowledgements}


\end{document}